\documentclass[%
 amsmath,amssymb,
 aip,
]{revtex4-2}

\usepackage{graphicx}
\usepackage{dcolumn}
\usepackage{bm}

\usepackage[utf8]{inputenc}
\usepackage[T1]{fontenc}
\usepackage{mathptmx}

\usepackage{color}
\usepackage[dvipsnames]{xcolor}

\newcommand{\bv}[1]{\mathbf{#1}} 
\newcommand{\dd}{\mathrm{d}}
\newcommand{\pp}[1]{\mathbf{#1}_{\perp}}

\begin{document}

The submitted manuscript has been created by UChicago Argonne, LLC, Operator of Argonne National Laboratory(“Argonne”). Argonne, a U.S. Department of Energy Office of Science laboratory, is operated under Contract No. DE-AC02-06CH11357. The U.S. Government retains for itself, and others acting on its behalf, a paid-up nonexclusive, irrevocable worldwide license in said article to reproduce, prepare derivative works, distribute copies to the public, and perform publicly and display publicly, by or on behalf of the Government.
\newpage


\title{High resolution functional imaging through Lorentz transmission electron microscopy and differentiable programming}

\author{Tao Zhou}%
\affiliation{ 
Nanoscience and Technology Division, Argonne National Laboratory, Lemont, IL 60439.
}%
\author{Mathew Cherukara}
\affiliation{ 
Advanced Photon Source, Argonne National Laboratory, Lemont, IL 60439.
}%
\author{Charudatta Phatak}
 \email{cd@anl.gov}
\affiliation{%
Materials Science Division, Argonne National Laboratory, Lemont, IL 60439.
}%

\date{\today}

\begin{abstract}
Lorentz transmission electron microscopy is a unique characterization technique that enables the simultaneous imaging of both the microstructure and functional properties of materials at high spatial resolution. The quantitative information such as magnetization and electric potentials is carried by the phase of the electron wave, and is lost during imaging. In order to understand the local interactions and develop structure-property relationships, it is necessary to retrieve the complete wavefunction of the electron wave, which requires solving for the phase shift of the electrons (phase retrieval). Here we have developed a method based on differentiable programming to solve the inverse problem of phase retrieval, using a series of defocused microscope images. We show that our method is robust and can outperform widely used \textit{transport of intensity equation} in terms of spatial resolution and accuracy of the retrieved phase under same electron dose conditions. Furthermore, our method shares the same basic structure as advanced machine learning algorithms, and is easily adaptable to various other forms of phase retrieval in electron microscopy.
\end{abstract}

\maketitle
\section{\label{sec:intro}Introduction} 
The design and development of new and improved materials requires a fundamental understanding of the structure-property relationship at the nanoscale. Vital to this understanding is the ability to quantify the local functional response of materials under \textit{operando} conditions. For example, imaging of the local magnetization under applied electromagnetic fields can give insight into novel magnetic proprieties emerging at the nanoscale \cite{chai2020,Jiang2019}. Similarly, quantification of the local electrostatic potential in solid electrolyte materials has proven crucial to the prediction of their charge transport behavior \cite{tavabi2011,swift2019,Xu2020}. Last but not least, the precise measurement of local strain is critical in many condensed matter systems such as semiconductors, ferroelectrics, and quantum materials \cite{Hong71,Whiteley2019}.  \\\indent
In transmission electron microscopy (TEM), the quantitative information about functional material properties is carried in the phase shift of the electrons \cite{Phatak2016}. The electron-sample interaction can be described as perturbations to the electron wavefunction given by $\psi(\mathbf{r}_{\perp})=a(\mathbf{r}_{\perp})e^{i\varphi(\mathbf{r_{\perp}})}$, where $\mathbf{r_{\perp}}$ is a radial vector along the electron propagation direction\cite{DeGraef2003IntroductionMicroscopy}. Elastic scattering of the electrons in the sample gives rise to variations in the amplitude of the wavefunction, $a(\mathbf{r}_{\perp})$, whereas the presence of electromagnetic potentials and strain field give rise to the variations in the phase of the wavefunction, $\varphi(\mathbf{r_{\perp}})$. The electromagnetic potentials here consist of two components namely (i) the electrostatic potential which is related to the local charge densities in the sample, and (ii) magnetic vector potential which is related to the magnetic spin texture of the sample. Depending on the imaging resolution, the electrostatic potential can be crystalline lattice potential (high resolution TEM) or charge accumulation due to electronic effects such as Mott-Schottky or 2D electron gas \cite{Hytch2011}. This phase information is however lost during image acquisition as the recorded intensity is merely the squared amplitude of the exit wavefunction. Quantitative evaluation of the functional properties thus necessitates solving for the phase of the electron wave. The process of obtaining phase information from measured intensities (known as phase retrieval) is the basis for a variety of coherent imaging techniques in x-ray and electron microscopy. In Lorentz TEM (LTEM), methods based on \textit{transport of intensity equation} (TIE) and off-axis electron holography are commonly used, each having its own merits and limitations \cite{Koch2010}. The TIE approach is experimentally easy to implement, but the result often suffers from low spatial resolution due to the extent of defocusing required. Off-axis holography on the other hand has high sensitivity, and high spatial resolution, but imposes stricter experimental conditions such as the need for a reference electron wave, and special electron-optical setup. \\\indent
Recent years have seen a tremendous increase in the application of machine-learning methods to determine the structure-property relationship in materials characterization. For electron microscopy, the primary focus has been on the analysis of high resolution images to determine atomic positions, or on the processing of large datasets for electron diffraction such as in 4D STEM\cite{Spurgeon2020}. There have been relatively fewer efforts in using such methods to obtain quantitative information about functional properties of materials. In this work, we have developed a method based on reverse-mode automatic differentiation (AD) to solve the inverse problem of phase retrieval in Lorentz transmission electron microscopy. AD has become the \textit{de facto} means of training a neural network \cite{Baydin2015AutomaticSurvey} thanks to the flexible interfaces developed and supported by large technology corporations \cite{Abadi2016TensorFlow:Systems, Paszke2019PyTorch:Library}. Phase retrieval with AD was first proposed \cite{Jurling2014ApplicationsAlgorithms} and later demonstrated with X-ray coherent diffraction imaging \cite{Nashed2017DistributedPtychography, Kandel2019UsingReconstruction}. By applying this method to LTEM, we show that we are able to retrieve phase at a higher spatial resolution and with higher accuracy as compared to conventional methods. \\\indent
\section{\label{sec:results}Results}
\subsection{\label{sec:r0}Phase retrieval using differentiable programming}
\begin{figure}[ht]
\includegraphics[width=0.9\textwidth]{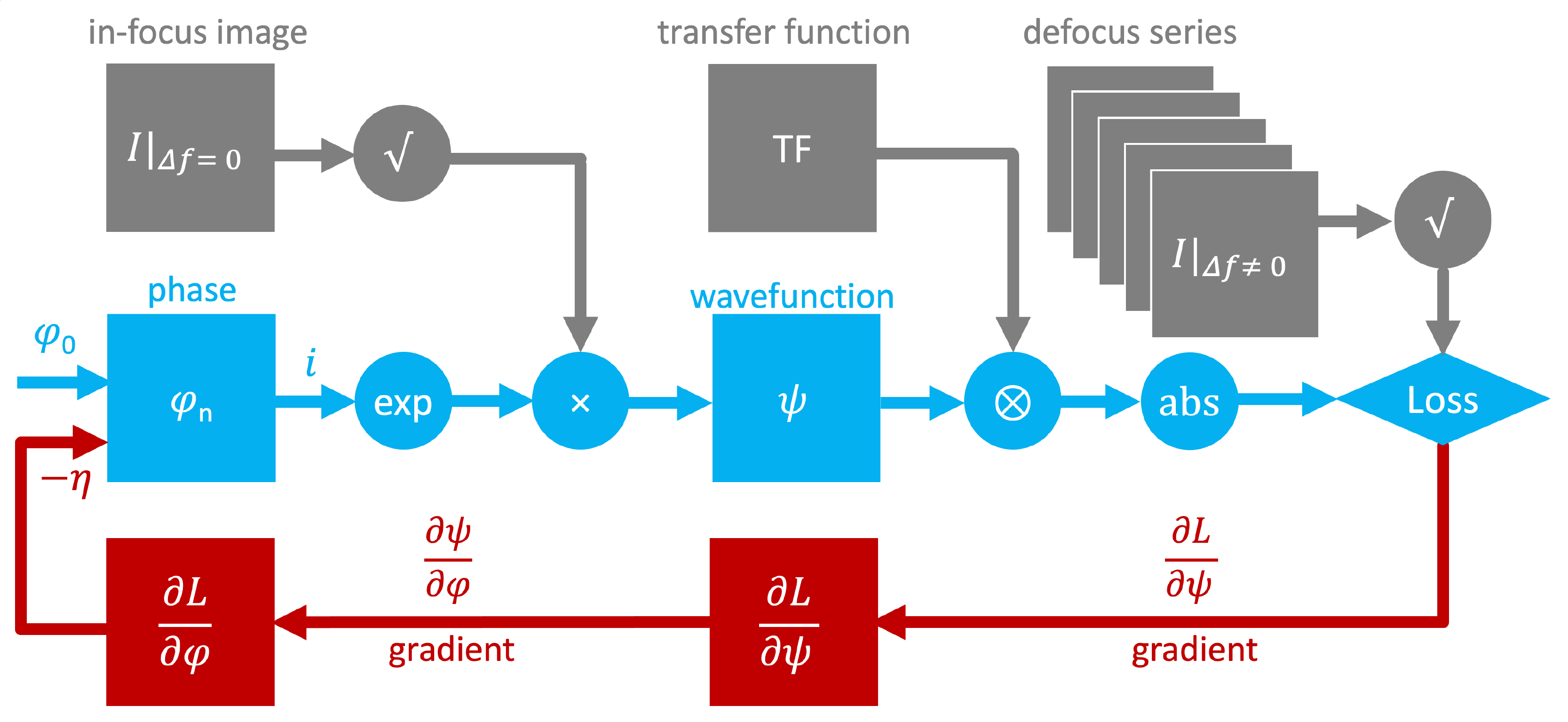}
\caption{\label{fig:one} Flow chart of phase retrieval in LTEM through differentiable programming. The experimental data and instrument parameters are color coded in grey. The imaging process that carries the phase information is color coded in blue while the path for back-propagation is color coded in red.}
\end{figure}
The process of applying AD for phase retrieval in LTEM is shown in Fig.~\ref{fig:one}. The amplitude $a(\pp{r})$ is fixed, and is taken as the square root of the intensity of the in-focus image. The starting guess $\varphi_0$ for the phase is set to a constant. For each iteration, the calculated data is computed as the exit wavefunction of the sample, $a(\pp{r})\text{exp}(i\varphi_n)$, convolved with the microscope transfer function (TF). The difference between the absolute of the calculated data and the square root of the measured intensity of the defocus series is used to compute the loss function. The concept of phase retrieval with AD is analogous to the training of a neural network (NN). In both cases, the gradients are calculated by back-propagating the loss through the network in what is known as reverse-mode automatic differentiation. The guess $\varphi_n$ is then updated iteratively in steps proportional to the negative of the gradients, and the phase retrieval process is considered complete if the improvement of the loss function is smaller than a pre-defined tolerance. For more detailed description, the reader is referred to the \textbf{Methods} section.\\\indent
\subsection{\label{sec:r1}Phase retrieval on image pairs of opposite defocus}
\begin{figure*}[hbt]
\includegraphics[width=0.9\textwidth]{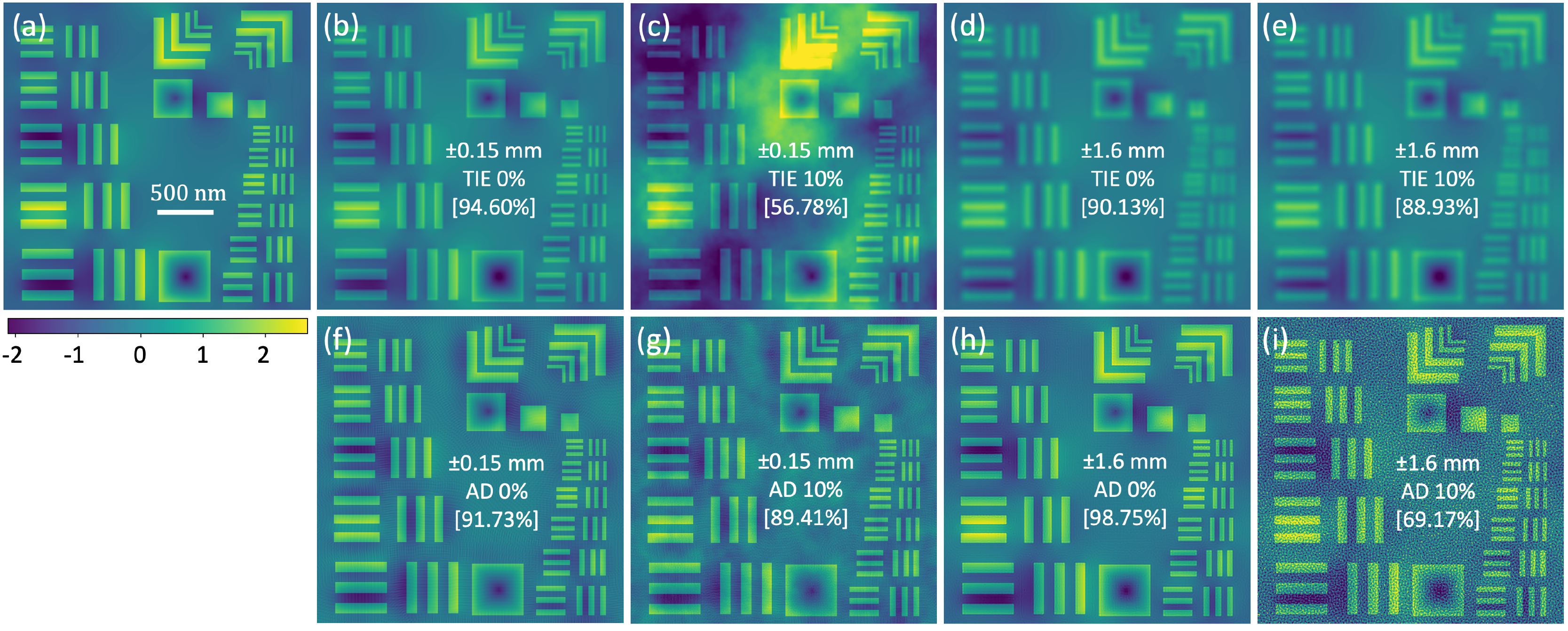}
\caption{\label{fig:two} (a) Ground truth for the phase. TIE retrieved phase using image pair with (b) 0\% and (c) 10\% of Gaussian noise at $\Delta f = \pm 0.15$ mm. TIE retrieved phase using image pair with (d) 0\% and (e) 10\% of Gaussian noise at $\Delta f = \pm 1.6$ mm. AD retrieved phase using image pair with (f) 0\% and (g) 10\% of Gaussian noise at $\Delta f = \pm 0.15$ mm. AD retrieved phase using image pair with (h) 0\% and (i) 10\% of Gaussian noise at $\Delta f = \pm 1.6$ mm. The numbers in the brackets denote the accuracy of retrieved phases.}
\end{figure*}

State-of-the-art TIE\cite{Phatak2016,ZUO2020106187} formalism uses the differentiation of two images of opposite defocus to approximate the longitudinal intensity derivatives ${\partial I}/{\partial z}$. The phase is then retrieved by solving the partial differential equation below using the intensity of the in-focus image, $I_0$,
\begin{equation}\label{eqn:tie}
\nabla\cdot [I_0\nabla\varphi]=-\frac{2\pi}{\lambda}\frac{\partial I}{\partial z}
\end{equation}\indent
The choice of this image pair has significant impact on the accuracy and spatial resolution of the TIE retrieved phase, as is illustrated in the upper part of Fig.~\ref{fig:two}. Fig.~\ref{fig:two}a shows the ground truth phase used for creating the simulated LTEM dataset (Fig.~\ref{fig:esone}). The knowledge of the ground truth is essential for evaluating the accuracy and spatial resolution of the retrieved phase, which is described in the \textbf{Methods} section. TIE retrieved phase using moderately defocused image pair ($\Delta f \approx \pm0.1$ mm) preserves a reasonable spatial resolution (Fig.~\ref{fig:two}b), but is less accurate and extremely susceptible to noise (Fig.~\ref{fig:two}c). TIE retrieved phase using strongly defocused image pair ($|\Delta f| > 1 $mm) is tolerant to noise (Fig.~\ref{fig:two}d and \ref{fig:two}e), but the result is blurry due to the much reduced spatial resolution\cite{McVitie2006}.\\\indent
We then performed AD phase retrieval on the same datasets. In the case of moderately defocused image pair, AD is not as accurate as TIE (Fig.~\ref{fig:two}f) under noise free conditions, but is able to retain the same level of accuracy in the presence of noise (Fig.~\ref{fig:two}g). In the case of strongly defocused image pair, AD vastly outperformed TIE under noise free conditions by simultaneously showing an extremely high level of accuracy and high spatial resolution (Fig.~\ref{fig:two}h). In the presence of noise, however, AD overfits to the noise, resulting in grainy images with low accuracy (Fig.~\ref{fig:two}i). Supplemental Figure \ref{fig:estwo}b shows the evolution of the accuracy. It can be seen that AD initially reached a higher level of accuracy (orange line) than TIE (orange rectangle), but did not converge to the truth despite a monotonically decreasing loss function (Supplemental Figure \ref{fig:estwo}c).\\\indent
\subsection{\label{sec:r2}AD phase retrieval on images of hybrid defocus}
\begin{figure*}[hbt]
\includegraphics[width=0.9\textwidth]{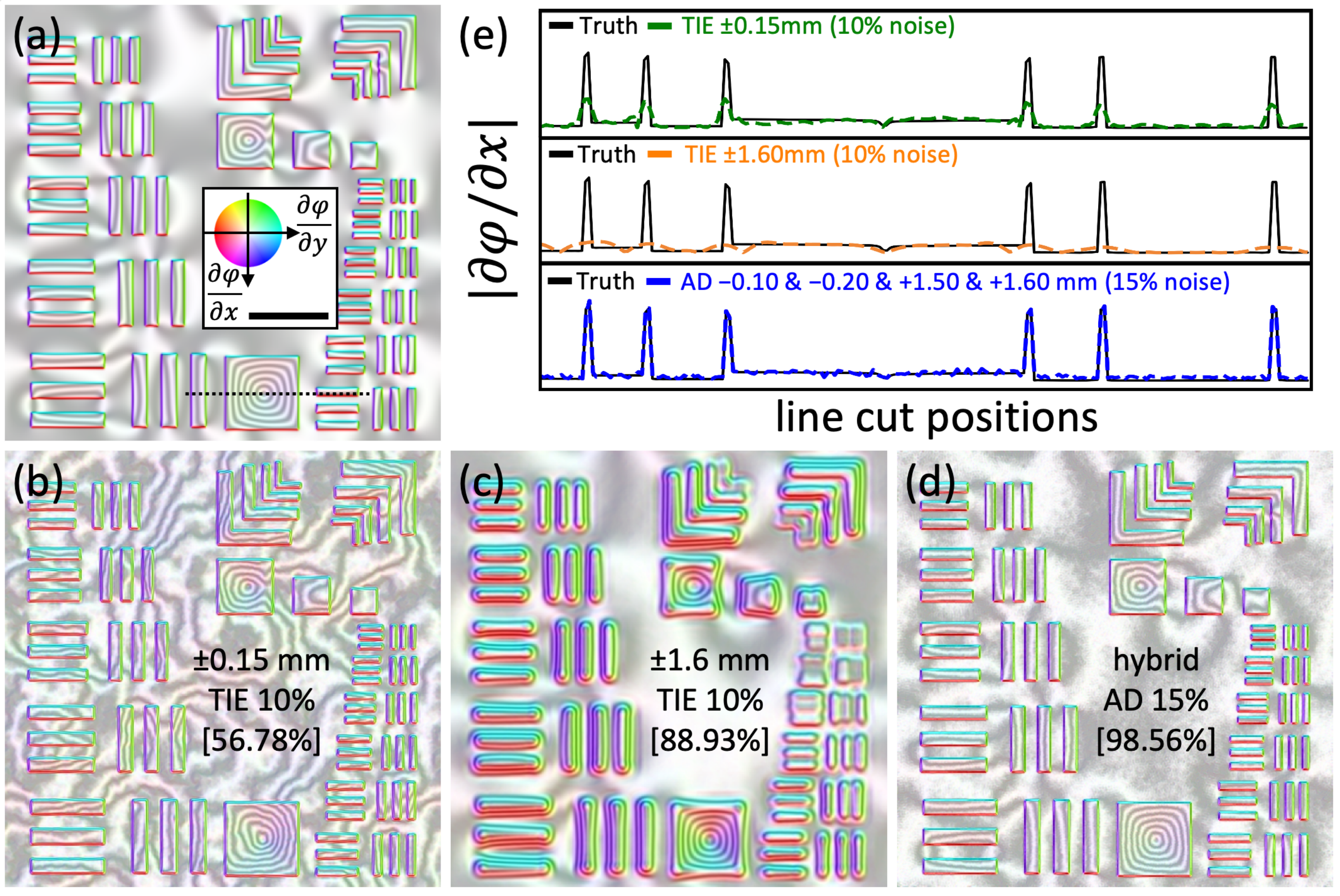}
\caption{\label{fig:three} (a) Ground truth for the phase gradient. Its direction is shown in false color as defined by the color wheel in the inset. Its magnitude is indicated by the gray contour which appears each time the phase wraps over one tenth of 2$\pi$. The scale bar is 500 nm. Gradient of TIE retrieved phase using image pair with 10\% Gaussian noise at (b) $\Delta f = \pm 0.15$ mm and (c) $\Delta f = \pm 1.6$ mm. (d) Gradient of AD retrieved phase using the 4 defocused images with 15\% Gaussian noise at $-0.1$, $-0.2$, $+1.5$ and $+1.6$ mm. (e) Detailed comparison with the ground truth respectively for data shown in (b), (c) and (d). The position at which the line profile was extracted is indicated by the dashed line in (a).}
\end{figure*}

The high accuracy and high spatial resolution achieved in Fig.~\ref{fig:two}h exemplifies the huge potential of AD as a viable phase retrieval method for LTEM, but its practical application is hampered by its instability in the presence of noise (Supplemental Figure \ref{fig:estwo}b). To circumvent this, we explore a strategy that leverages the flexibility of the AD method to work with multiple images at different defocus conditions. We choose to show the gradient of the phase to highlight the effectiveness of our strategy. Compared to the phase itself, the gradient of the phase is sensitive to the direction of magnetic induction or local electric fields as well as to the edges of nanostructures due to change in thickness. As a result, the phase gradient gives a better assessment of the spatial resolution and accuracy of the retrieved phase. Fig.~\ref{fig:three}a shows the ground truth for the gradient of the phase, with its direction and its magnitude respectively indicated by false color and grayscale contour. With 10\% of Gaussian noise, the magnitude of the TIE retrieved phase gradient at moderate defocus is severely distorted at the center of the nanostructures (Fig.~\ref{fig:three}b), which is reflective of the low accuracy of 56.78\%. Under strongly defocused conditions, the TIE retrieved phase gradient is distorted near the edge of the nanostructures (Fig.~\ref{fig:three}c), as a result of the low spatial resolution. \\\indent
Next, we demonstrate that high spatial resolution, high accuracy and tolerance to noise can be simultaneously achieved in the AD formalism. We recall that while AD seems to guarantee a high spatial resolution regardless of the focusing conditions, extremely high accuracy was only observed with 
strongly defocused image pairs while tolerance to noise was only observed with moderately defocused image pairs. One advantage of the AD phase retrieval process is that it does not require the images to be pairs of opposite defocus. Indeed, pairing one image of moderate defocus with one of strong defocus effectively stabilized the accuracy evolution in the presence of 10\% Gaussian noise (Fig.\ref{fig:esthree}b, red line). The best result (Fig.~\ref{fig:three}d) was obtained by combining two images of moderate defocus with two images of strong defocus, where an accuracy of 98.56\% was reached (Fig.\ref{fig:esthree}b, blue line), about 10\% higher than the best value achievable with the TIE method. To ensure a fair comparison, the noise level of the four images was raised to 15\%. This is to account for the factor of $\sqrt{2}$ shot noise variation when the exposure time of each image is halved and the total exposure time remains unchanged. The much improved accuracy in the final retrieved phase is thus understood as due entirely to a stronger numerical constraint. Fig.~\ref{fig:three}e shows the line profiles of the retrieved phase gradient across multiple nanostructures. It can be seen that AD with hybrid defocus images is the only method capable of reproducing faithfully the sharp variation of the phase gradient at the edge of the nanostructures.\\\indent
\subsection{\label{sec:r3}Application on experimental data}
\begin{figure}[t]
\includegraphics[width=0.9\textwidth]{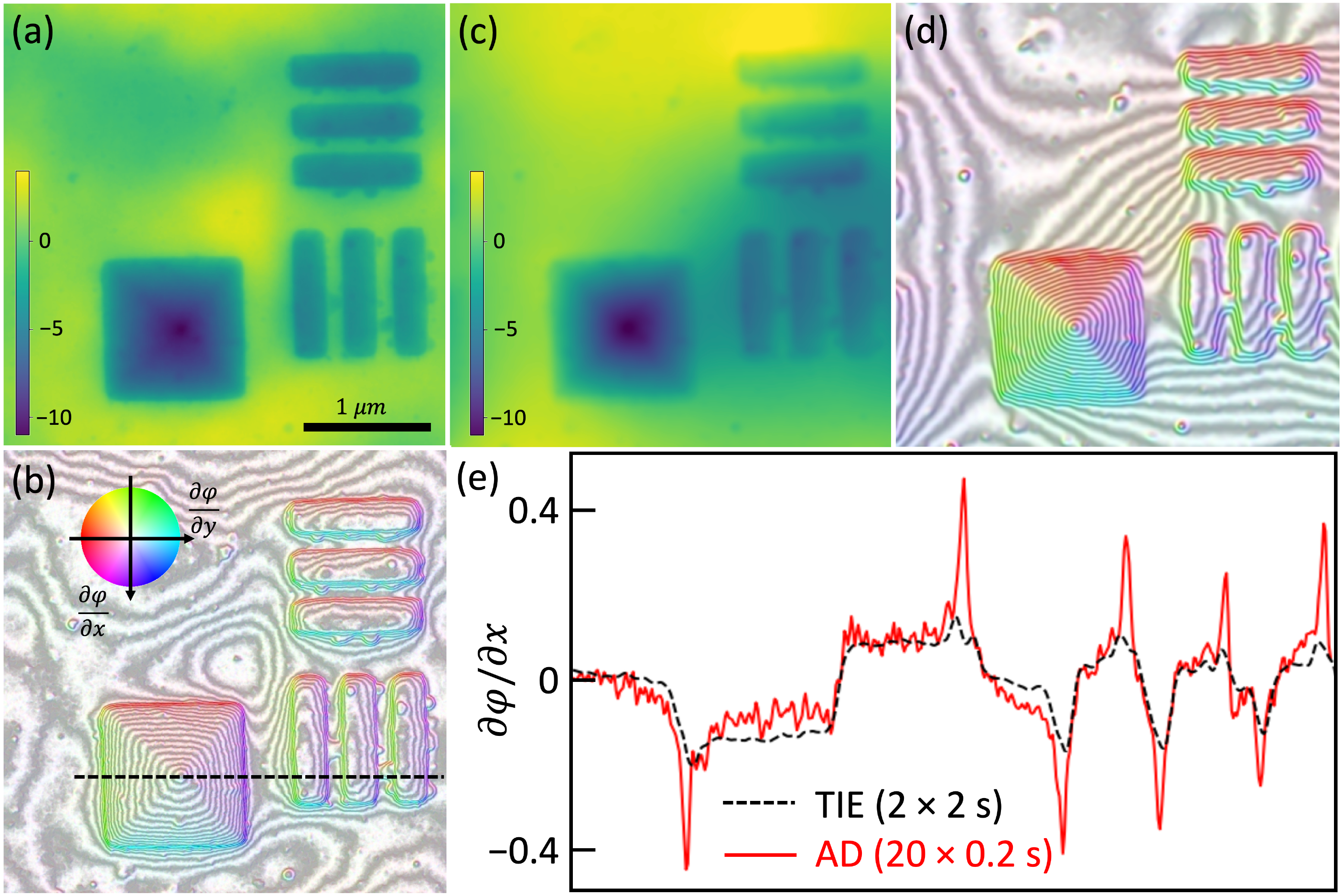}
\caption{\label{fig:four} (a) AD retrieved phase and (b) its gradient on images at defocus spanned between $\Delta f = \pm 1.44$ mm, with 4 s of total exposure time. (c) TIE retrieved phase and (d) its gradient on images at defocus $\Delta f = \pm 1.44$ mm, with 4 s of total exposure time. (e) shows line profile of the phase gradient. The position at which the line profile was extracted is indicated by the dashed line in (b).}
\end{figure}

Finally, we demonstrate the viability of the proposed phase retrieval strategy on experimental LTEM images. The imaging conditions and information about the sample can be found in the \textbf{Methods} section. We have specifically chosen, for the purpose of verification, nanostructures with known magnetic configuration. Three different scenarios were tested, with 2, 4 and 20 images taken at various defocus conditions. The total counting time was 4 s in all three cases, to ensure a fair comparison under the same electron dose conditions. Similar to what was concluded with the simulated dataset, the convergence of the AD method improves with increasing number of images, with the best result obtained for a series of 20 images with hybrid defocus. Fig.~\ref{fig:four}a and \ref{fig:four}b show respectively the AD retrieved phase and its gradient on the 20 defocus images spanned between $\Delta f = \pm 1.44$ mm. The exposure time was 0.2 s per image. Fig.~\ref{fig:four}c and \ref{fig:four}d show in comparison the TIE retrieved phase and its gradient at $\Delta f = \pm 1.44$ mm, with 2 s of exposure time per image. As expected, AD retrieved phase (Fig.~\ref{fig:four}a) appears to be sharper than the TIE retrieved one (Fig.~\ref{fig:four}c) thanks to its inherent high spatial resolution. Line profiles were extracted across the center of the square-shaped nanostructure. Once again, AD (Fig.~\ref{fig:four}e, red line) outperformed TIE (black line) in retrieving the sharp variations of the phase gradient expected at the edges of the nanostructures. Elsewhere on the
extracted line profiles, the two methods agree extremely well with each other, indicating that the accuracy of the AD method on the experimental data is at least equal to, if not better than that of TIE, under the same electron dose conditions.\\\indent
\section{\label{sec:Discussion}Discussion}
\begin{figure}[t]
\includegraphics[width=0.9\textwidth]{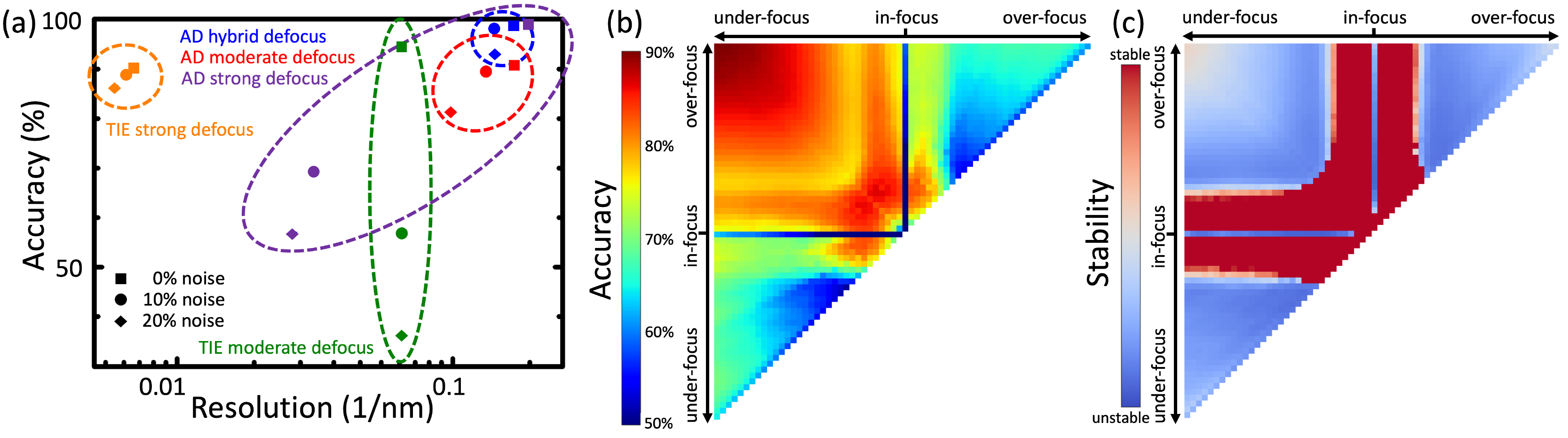}
\caption{\label{fig:five} (a) Accuracy versus resolution plot for the TIE and the AD methods. The values are calculated from phase retrieved on the simulated dataset with various noise levels. Moderate defocus refers to $\Delta f = \pm 0.15$ mm while strong defocus refers to $\Delta f = \pm 1.6$ mm. The hybrid defocus uses 4 defocused images at -0.1, -0.2, +1.5 and +1.6 mm. (b) Accuracy of the AD retrieved phase using 2 images of hybrid defocus, for data with 10\% of Gaussian noise. The defocus values cover the entire parameter space from $\Delta f = + 1.6$ mm (under-focus) to $\Delta f = - 1.6$ mm (over-focus). The values shown are the final accuracy of the retrieved phase after 500 iterations. (c) Stability of the AD retrieved phase covering the same parameter space as (b). The more "unstable" an image pair is, the earlier in the iterations does the AD retrieved phase diverges from the ground truth. The maximum value corresponds to those image pairs that continue to converge to the ground truth after 500 iterations.}
\end{figure}
In this work, we have demonstrated a new method based on automatic differentiation for the phase retrieval in Lorentz TEM. The strengths and weaknesses of our proposed method against the conventional TIE approach are summarized in Fig.~\ref{fig:five}a, based on results presented in Fig.~\ref{fig:two} and Fig.~\ref{fig:esfour}. TIE in theory requires a pair of images, moderately defocused in the opposite directions. While that works well with ideal data, the accuracy of the retrieved phase drops quickly in the presence of noise (green symbols). The tolerance to noise can be greatly enhanced by increasing the defocus distance of the image pair (orange symbols). This can be easily understood by looking back at Eq.~\ref{eqn:tie}, as any statistical fluctuations in the intensity is attenuated by a factor proportional to the defocus distance. The use of strongly defocused images is not without its problems. TIE is strictly valid only in the small defocus limit where the microscope transfer function remains linear for the given spatial frequency resolution\cite{BAJT200067, DeGraef2001}. Therefore, the strong defocusing reduces the achievable spatial resolution, resulting in blurriness in the retrieved phase (Fig.~\ref{fig:two}d).\\\indent
We then demonstrated the use of AD phase retrieval on the same set of data. We note that good spatial resolution is always guaranteed in the AD formalism regardless of the defocus distance of the image pair. But a new dilemma emerges. The accuracy of the retrieved phase is limited to about 90\% when using moderately defocused images (red symbols). This can be understood by the fact that the transfer function for moderate defocus values does not have much variations to carry enough phase information into image intensity. Extremely high accuracy can be obtained by using pairs of strongly defocused images (purple symbols). However, the achieved high accuracy was limited to ideal data only, as AD tends to overfit to the presence of noise, resulting in divergence from the ground truth in noisy images. We solve this dilemma by leveraging the flexibility of the AD method to work with multiple images of mixed defocus conditions. In the most simple scenario, this involves pairing two images taken at different defocus distances. We have mapped out the entire parameter space to determine the achievable accuracy and stability (defined as the ability to stay converged to the ground truth in the presence of noise) in this scenario. The result is shown respectively in Fig.~\ref{fig:five}b and c. It can be seen that while the maximum accuracy is obtained by pairing one strongly under-focused image with one strongly over-focused one (equivalent to having a strongly defocused image pair), the stability of the AD approach is not very high under these conditions. Instead, we can identify a sweet spot for high accuracy and high stability where a strongly defocused image is paired with a moderately defocused one (referred to as hybrid defocus).\\\indent
We further demonstrated that, under the same electron dose condition, the accuracy and stability of AD phase retrieval improves with the number of images. For simulated data, the accuracy of the retrieved phase is 72.53\% for 2 images of 10\% Gaussian noise, and 98.56\% for 4 images of 15\% noise (Fig.~\ref{fig:five}, blue symbols). Additional tests on simulated data shows that the noisier the data is, the larger the number of images is required to maintain the stability (Fig.~\ref{fig:esfive}). The same was observed in the application of AD phase retrieval on experimental LTEM images, but for a slightly different reason. The retrieved phase appears to be more accurate when 20 images of hybrid defocus were used, as compared to 2 or 4 images of equal total exposure time. We believe that the use of a large number of images averages out the variations introduced by uncertainties in experimental parameters such as sample misalignment, leading to a highly accurate retrieved phase in our case.\\\indent
The most significant advantage of our proposed method, as compared to the conventional TIE approach, is its ability to retrieve phase with simultaneous high spatial resolution and high accuracy. This is best illustrated on the line profiles of the gradient of the phase in both the simulated (Fig.~\ref{fig:three}e) and experimental (Fig.~\ref{fig:four}e) dataset. Compared to off-axis electron holography, it offers high resolution phase information over a large field of view, without the additional experimental complications. This is particularly important for applications such as determining interfacial electrostatic potentials in materials where it is critical to sample a statistically significant number of interfaces. The main limitation of our method is its requirement of \textit{a priori} knowledge of the microscope transfer function. Some of the parameters in the TF can be accurately determined in an aberration-corrected Lorentz TEM, such as spherical aberration and defocus distance. Other parameters like beam coherence are experimentally harder to evaluate, in which case, we recommend extending the AD approach to also retrieve these parameters along with the sample phase. It is in theory possible to recover the true amplitude as well, provided that sufficient amount of data is taken to ensure a strong numerical constraint. This would enable the reconstruction of the complete electron wavefunction, which can be adapted for high-resolution TEM imaging at the atomic scale.  Finally, thanks to the simplicity of its implementation, the method can be easily incorporated with other iterative reconstruction algorithms such as electron beam tomography to determine three-dimensional electromagnetic fields.\cite{venkatakrishnan2014model, fanelli2008electron, Mohan2018}. 
\\\indent
\section{\label{sec:Methods}Methods}
\subsection{\label{sec:Methods1}Gradient descent optimization with automatic differentiation}
We use the Adam optimizer \cite{Kingma2015Adam:Optimization} implemented in Google's Tensorflow package\cite{tensorflow2015-whitepaper} for the gradient descent optimization. The initial learning rate is set to 1. Cyclic learning rate \cite{Smith2015} has occasionally been used but does not show additional improvement on the maximum achievable accuracy. Before computing the amplitude, the in-focus image was denoised with a total variation (TV) filter. For simulated data, the weight of the TV filter is set to the level of the Gaussian noise. For experimental data, it is set to the standard deviation measured on the background area. No denoising process was performed on any of the defocus images. The starting guess for the phase is 0.5 everywhere. Mean Squared Error was used as the loss function. The phase retrieval process is run on a remote Nvidia Tesla K80 GPU hosted on Google's Colaboratory. The amount of time per iteration varies depending on the number of defocus images used, but is typically 2 min per 5000 iterations.\\\indent
\subsection{\label{sec:Methods2}Accuracy and spatial resolution of the retrieved phase}
The use of the simulated datasets allowed us to evaluate the accuracy of the retrieved phase, calculated as the correlation between the ground truth phase $\varphi_\text{tru}$ (Fig.~\ref{fig:two}a) and the retrieved phase $\varphi_\text{ret}$. The sum $\Sigma$ runs over each pixel of the 2D image.
\begin{equation}\label{eqn:ab_correlation}
    \Sigma|\varphi_\text{ret}\varphi_\text{tru}|/\sqrt{ \Sigma|\varphi_\text{ret}\varphi_\text{ret}| \Sigma|\varphi_\text{tru}\varphi_\text{tru}|}
\end{equation}\\\indent
The spatial resolution of the retrieved phase is estimated by fitting a Gaussian function to the sharp variation of the absolute of the phase gradient at the edges of the nanostructures (Fig.~\ref{fig:four}e). The spatial resolution is then taken as the FWHM of the Gaussian distribution. Because an error function can be considered as twice the integral of a normalized Gaussian function. This is equivalent to fitting an error function to the sharp variation of the phase at the same edges (Fig.~\ref{fig:estwo}d and \ref{fig:estwo}e)
\subsection{\label{sec:Methods3}Simulated and experimental dataset}
Simulated LTEM images were computed for magnetic nanostructures using parameters for Permalloy (Fig.~\ref{fig:esone}). For a given noise level, a total of 65 images were produced with defocus values spanned evenly between $\Delta f =\pm 1.6$ mm. The magnetization of the nanostructures was determined using OOMMF micromagnetic simulations \cite{OOMMF}, from which the electron phase shift was calculated using the following equation
\begin{equation}\label{eqn:ab_phaseshift}
        \varphi_t(\pp{r}) 
        = \sigma\int V(\pp{r},z)~\dd z + \frac{e}{\hbar}\int\bv{A}(\pp{r},z)~\dd z;
\end{equation}
where $\sigma$ is the interaction constant for a TEM and depends on the accelerating voltage ($\sigma = 0.00728$ V.nm$^{-1}$ for $200$ kV electrons), $\hbar$ is the reduced Planck's constant and the integration is carried out along the direction of propagation of the electrons. The simulated pattern was a slightly modified version of the 1951 USAF resolution test standard. Each LTEM image is composed of 512$\times$512 pixels of 5$\times$5 nm. Gaussian noise was added to the images after they are generated. The level of the Gaussian noise refers to the standard deviation of the distribution.\\\indent
Experimental images (512$\times$512 pixels) were taken using an aberration-corrected JEOL 2100F Lorentz TEM operating at 200 kV. The sample consists of $10$ nm thick of  Permalloy magnetic nanostructures, sputter-deposited on a TEM grid and patterned with e-beam lithography. The pixel size was 6.9 nm.\\\indent

\bibliography{reference,cd_refs}

\begin{acknowledgments}
This work was supported by the U.S. Department of Energy, Office of Science, Basic Energy Sciences, Materials Sciences and Engineering Division. This work was performed (T.Z, M.C), in part, at the Center for Nanoscale Materials and the Advanced Photon Source, both U.S. Department of Energy Office of Science User Facilities, and supported by the U.S. Department of Energy, Office of Science, under Contract No. DE-AC02-06CH11357. We acknowledge Martin V. Holt for the insightful discussions and guidance.
\end{acknowledgments}

\section*{Author contributions}
T.Z and M.C implemented the forward model using differentiable programming in Google's Tensorflow package. C.P generated the simulated as well as the experimental images. All authors discussed the results and contributed to the manuscript.

 \section*{Competing interests} 
 The authors declare that they have no competing financial interests.
 
 \section*{Data Availability} 
The data that support the findings of this study are available from the corresponding author upon reasonable request.

\newpage
\section{Supplementary Material}

\renewcommand{\figurename}{Supplemental Figure}
\renewcommand\thefigure{S\arabic{figure}}
\setcounter{figure}{0}  

\begin{figure*}[hbt]
\includegraphics[width=0.9\textwidth]{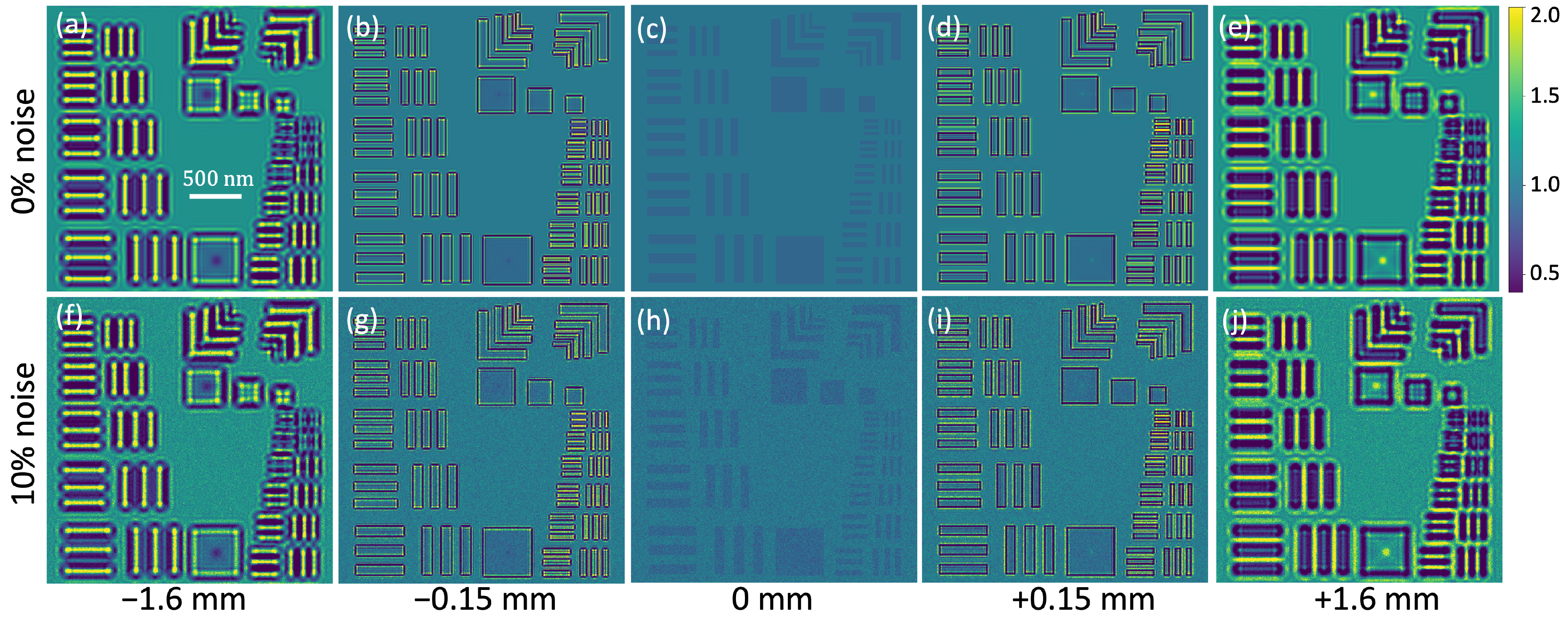}
\caption{\label{fig:esone} Simulated LTEM images. Simulated noise-free LTEM images at $\Delta f =$ (a) -1.6 mm, (b) -0.15 mm, (c) 0 mm, (d) +0.15 mm and (e) +1.6 mm. Simulated LTEM images with 10\% of Gaussian noise at $\Delta f =$ (f) -1.6 mm, (g) -0.15 mm, (h) 0 mm, (i) +0.15 mm and (j) +1.6 mm.}
\end{figure*}

\begin{figure*}[hbt]
\includegraphics[width=0.9\textwidth]{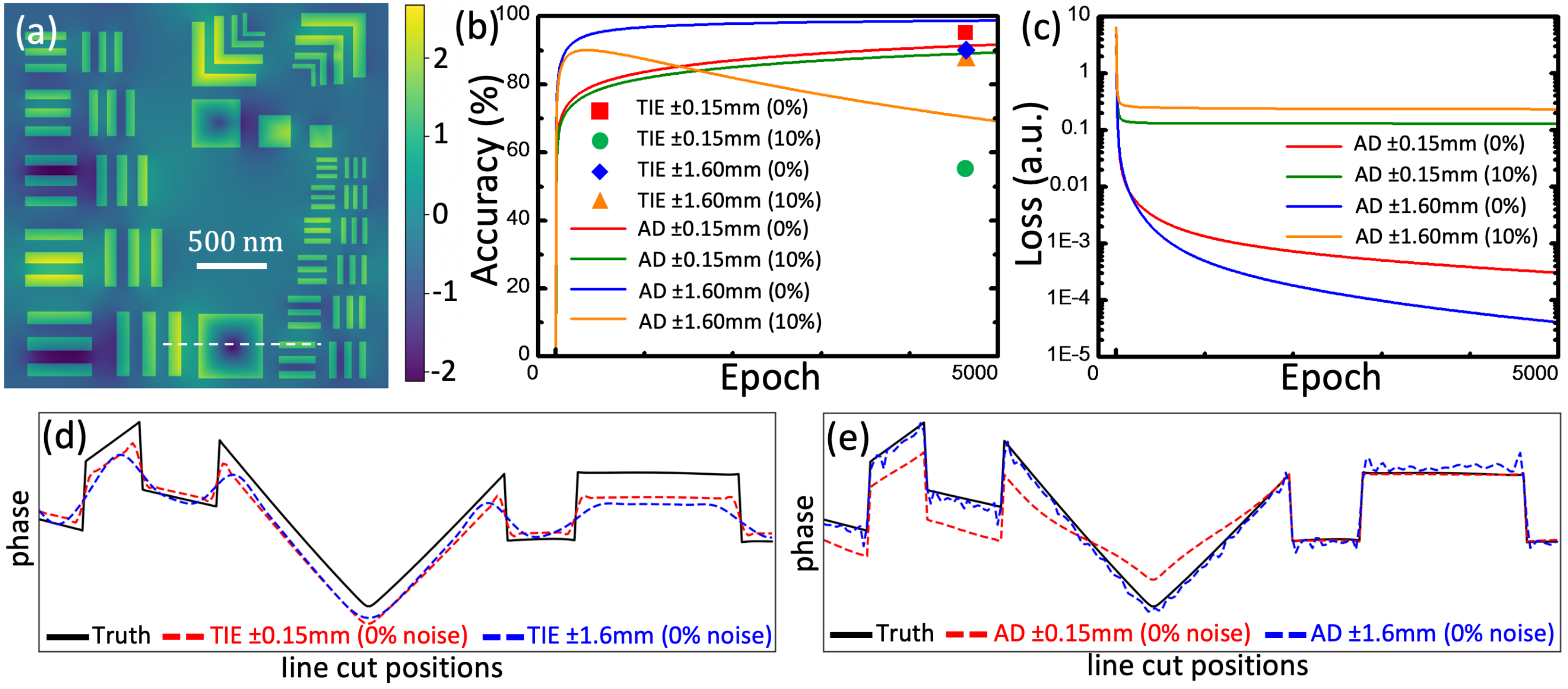}
\caption{\label{fig:estwo} Comparison between TIE and AD retrieved phases. (a) Ground truth for the phase. (b) Evolution of the accuracy during 5000 iterations of AD phase retrieval (learning rate = 1) in logarithmic scale. The accuracies for the TIE retrieved phases are also shown for comparison. (c) Evolution of the loss function during the same AD phase retrieval process. (d) and (e) show respectively line profiles of the TIE and AD retrieved phase. The position at which the line profile was extracted is indicated by the dashed line in (a).}
\end{figure*}

\begin{figure*}[hbt]
\includegraphics[width=0.9\textwidth]{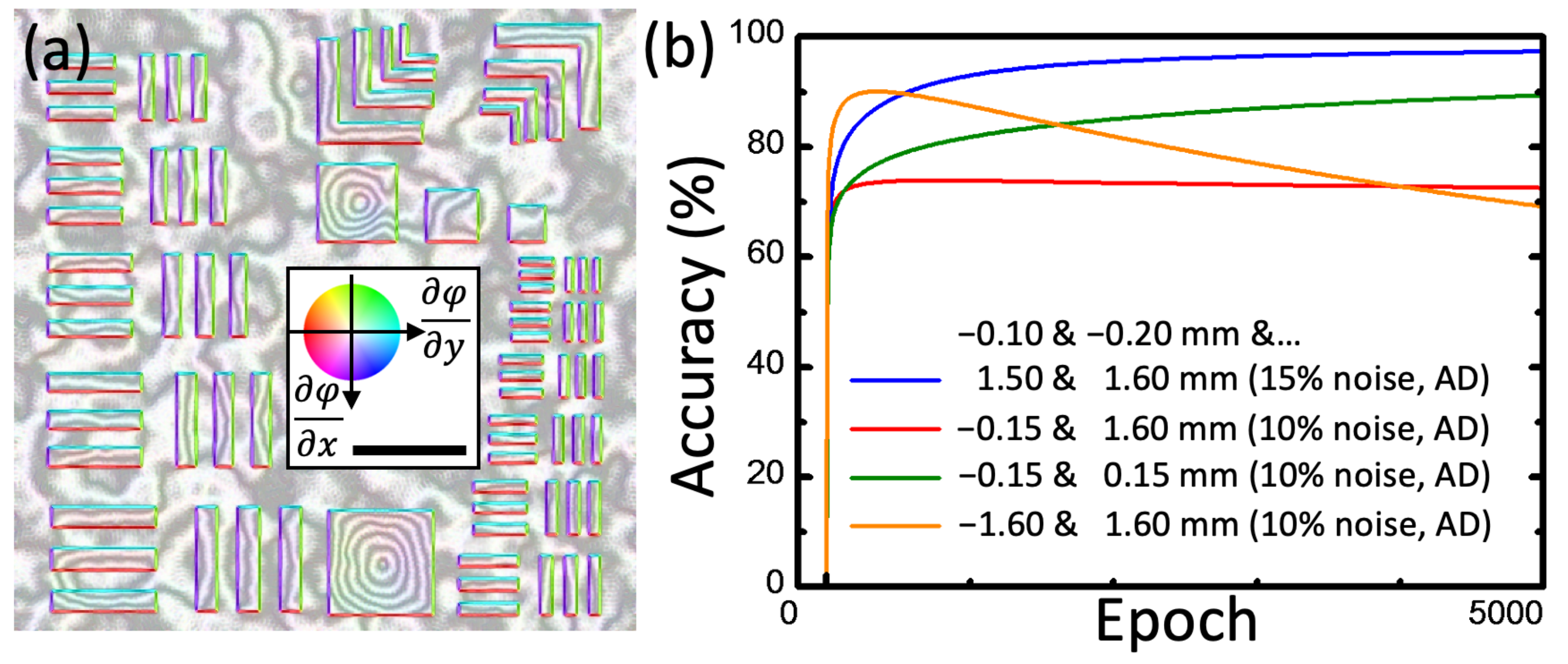}
\caption{\label{fig:esthree} AD retrieved phase with hybrid defocus. (a) AD retrieved phase using the 2 defocused images with 10\% Gaussian noise at $\Delta f = -0.15$ and $+1.6$ mm. (b) Evolution of the accuracy during 5000 iterations of AD phase retrieval (learning rate = 1) in logarithmic scale.}
\end{figure*}

\begin{figure*}[hbt]
\includegraphics[width=0.9\textwidth]{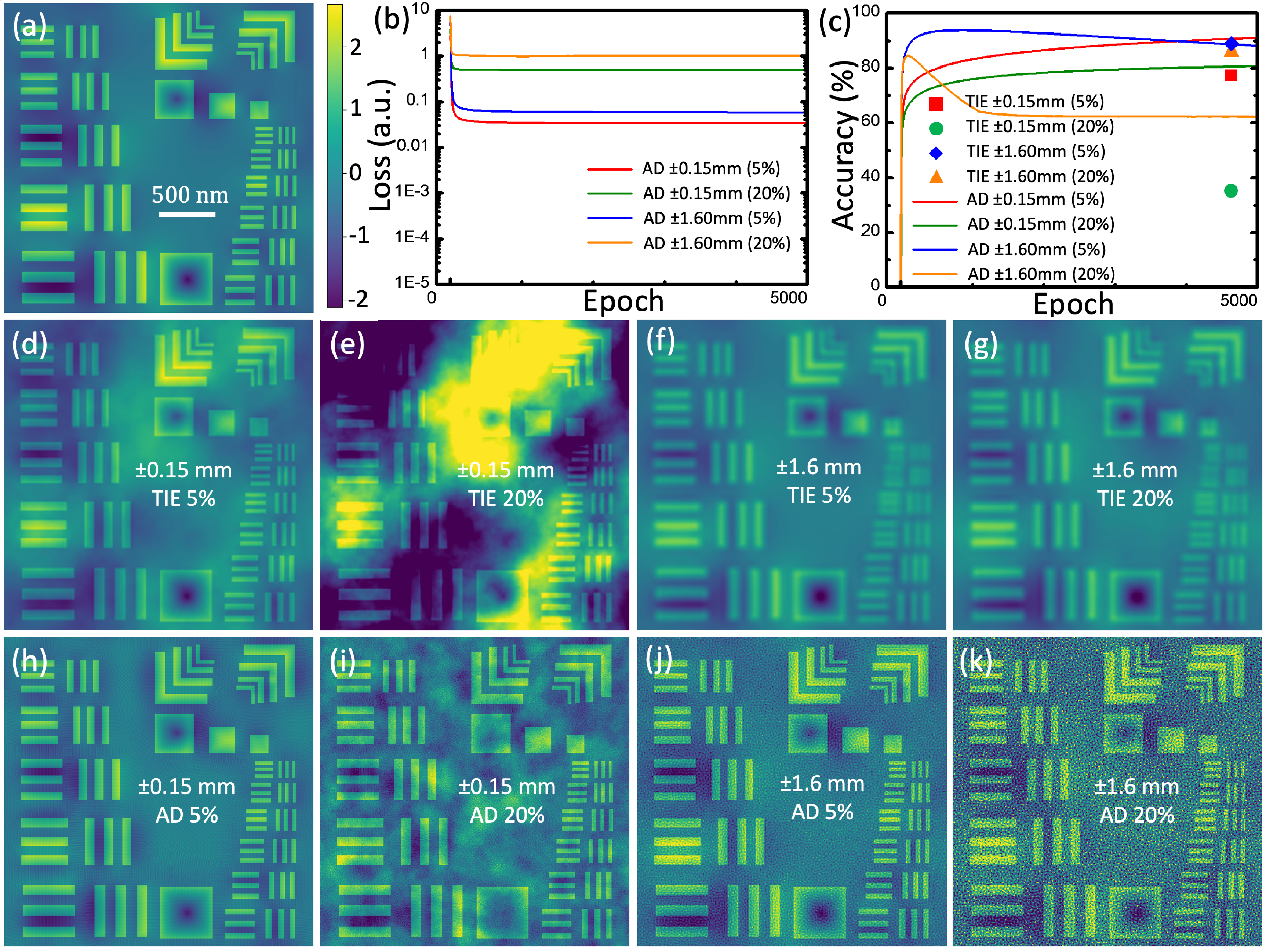}
\caption{\label{fig:esfour} (a) Ground truth for the phase. (b) Evolution of the loss function during 5000 iterations of AD phase retrieval (learning rate = 1) in logarithmic scale. (c) Evolution of the accuracy during the same AD phase retrieval process. The accuracies for the TIE retrieved phases are also shown for comparison. TIE retrieved phase using image pair with (d) 5\% and (e) 20\% of Gaussian noise at $\Delta f = \pm 0.15$ mm. TIE retrieved phase using image pair with (f) 5\% and (g) 20\% of Gaussian noise at $\Delta f = \pm 1.6$ mm. AD retrieved phase using image pair with (h) 5\% and (i) 20\% of Gaussian noise at $\Delta f = \pm 0.15$ mm. AD retrieved phase using image pair with (j) 5\% and (k) 20\% of Gaussian noise at $\Delta f = \pm 1.6$ mm.}
\end{figure*}

\begin{figure*}[hbt]
\includegraphics[width=0.9\textwidth]{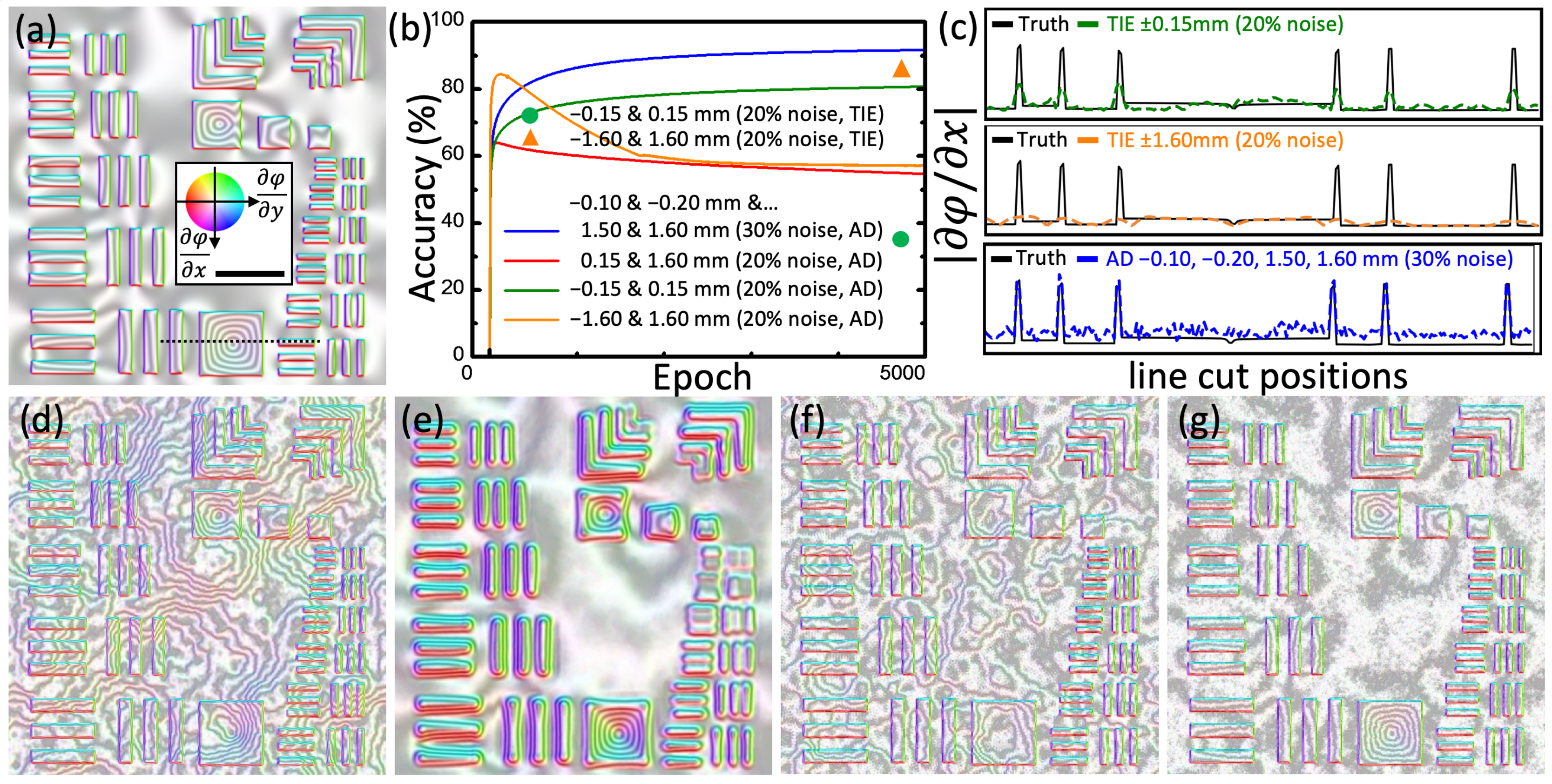}
\caption{\label{fig:esfive} (a) Ground truth for the phase gradient. Its direction is shown in false color as defined by the color wheel in the inset. Its magnitude is indicated by the gray contour which appears each time the phase wraps over one tenth of 2$\pi$. The scale bar is 500 nm. (b) Evolution of the accuracy during 5000 iterations of AD phase retrieval (learning rate = 1). The accuracies for the TIE retrieved phases are also shown for comparison. (c) Detailed comparison with the ground truth respectively for data shown in (d), (e) and (g). The position at which the line profile was extracted is indicated by the dashed line in (a). Gradient of TIE retrieved phase using image pair with 20\% Gaussian noise at (d) $\Delta f = \pm 0.15$ mm and (e) $\Delta f = \pm 1.6$ mm. (f) AD retrieved phase using the 2 defocused images with 20\% Gaussian noise at $-0.15$ and +1.6 mm. (g) AD retrieved phase using the 4 defocused images with 30\% Gaussian noise at $-0.1$, $-0.2$, $+1.5$ and $+1.6$ mm. The increased noise level is to account for the factor of $\sqrt{2}$ shot noise variation when the exposure time of each image is halved and the total exposure time remains unchanged.}
\end{figure*}

\end{document}